# Suppression of Aggregation in Natural-Semiflexible/Flexible Polyanion Mixtures, and Direct Check of the OSF Model using SANS

F. Bonnet,[1] R. Schweins,[1,*] F. Boué,[2] and E. Buhler [3,*]

[1]*Institut Laue Langevin, Large Scale Structures Group, 6 rue Jules Horowitz, BP 156, 38042 Grenoble Cedex 9, France*
[2]*Laboratoire Léon Brillouin, CEA-CNRS, CE Saclay, 91191 Gif/Yvette, France*
[3]*Matière et Systèmes Complexes, UMR CNRS 7057, Université Paris 7-Denis Diderot, Bâtiment Condorcet, CC 7056, 75205 Paris Cedex 13, France*

Aggregation and other interactions are suppressed for a biological semiflexible polyelectrolyte, hyaluronan (HA), when it is embedded in a mixture with another negatively charged and flexible polyelectrolyte chain, sodium polystyrene sulfonate. We see directly HA only in the mixture using Small-Angle Neutron Scattering, isotopic labelling and contrast matching. At low ionic strength, for which aggregation is usually seen for pure HA solutions, an unambiguous set of experimental results shows that we neither observe HA aggregation nor a polyelectrolyte peak (observed for solutions of single species); instead we observe a wormlike chain behaviour characteristic of single chain with a variation of the persistence length with the square of the Debye screening length, $L_e \sim \kappa^{-2}$, as formerly predicted by Odijk and not yet observed on a polymer chain.

In biological systems, polyelectrolytes [1-4] are present at each step of cell life. Proteins, globular and water-soluble, or membranar and amphiphilic, with compact regions, can be regrouped in a first ensemble; while DNA, RNA, forming a second ensemble, are closer to string-like chains made of fewer repeating units [2]. We focus on this side, with polymers of even simpler composition but important in vegetal and animal life, like polysaccharides. Hyaluronan (HA), which is studied here, is a polysaccharide made of a single negatively charged repeating unit. It is alike the polyelectrolytes considered by theory [1], thus making it a good model for physics.

HA is also a useful model for biology: beyond its well known role in articulations, it is, on a more general basis, found in extra-cellular matrix, in soft connective tissues as in cartilage, where it is forming complexes with proteins, glycoproteins and/or other electrostatically charged species. It has also been assigned various physiological functions in the intercellular matrix, e.g., in water and plasma protein homeostasis [3,4]. The polymer is assumed to play a role in mitosis, as HA production increases in proliferating cells. Interaction between HA and extra-cellular polysaccharide has been connected with locomotion and cell migration, which increases its interest in cancer research [3,4].

As a consequence of their large contribution to life, many different bio-polyelectrolytes are bound to interact. A first situation is related to positive-negative charge interactions: this is observed for interaction between polyelectrolyte chains and globular proteins: protein and polyelectrolyte chains also interact in the second case of global charges of the same sign. Here, we focus on a third case implying only chains, but of two different species: HA and a flexible polyelectrolyte. We can regard it starting from the case of polyelectrolyte solutions of a single chain species. Even for the more simple system of a salt-free solution of linear polyelectrolytes, much remains to understand. Radiation scattering gives access to a pertinent and best explained feature, the structure factor maximum ("polyelectrolyte peak") signalling a privileged repulsion distance between chains; this peak broadens and moves to higher scattering vectors q as the concentration is increased [5]. On the contrary, a major debatable point concerns the aggregation behaviour observed at small q and correlated to the slow mode measured by dynamic light scattering [6,7]. Indeed, it is surprising that two macromolecules with like charge **attract** while they repel at shorter distances. This occurs reproductively on different charged biopolymers, including DNA, proteins and polysaccharides when ionic strength, I, is low [6,7]. During the last decade, there has been exciting debates in the "polyelectrolyte community" to elucidate this phenomenon. In particular, a well-accepted model developed by Ray and Manning and based on an attractive force between polyions mediated by the sharing of condensed counterions is generally used to interpret these experimental facts [8]. Beyond its peculiarity, this low-q aggregation signal has the disadvantage of masking a length domain necessary for any determination of the chain persistence length, to which we come now.

Indeed, another question concerns the rigidity induced by inter like-charged monomer repulsion; this is quantified by an "electrostatic persistence length", $L_e$ that depends on I, which is due to counterions as well as external salt. There has been some theoretical debate about whether $L_e$ varies like $\kappa^{-1}$ or $\kappa^{-2}$, where $\kappa^{-1}$ ($\sim I^{-1/2}$) represents the Debye length [1]. These theories describe the polyelectrolyte as a "wormlike" chain with a total persistence length, $L_T = L_0 + L_e$, where $L_0$ is the intrinsic persistence length of the uncharged polymer. For semi-rigid polyelectrolytes, with $L_0$ easily larger than $L_e$, Odijk [9], and Skolnick and Fixman [10] found:

$$L_e = \frac{l_B}{4\kappa^2 b^2} \quad \text{for } l_B/b<1 \text{ (OSF relation, valid if } \kappa L_T \gg 1) \qquad (1)$$

The Debye length is $\kappa^{-1} = (4\pi l_B c_f)^{-1/2}$ (the Bjerrum length $l_B = 7.13$ Å in water), $c_f = c + 2c_s$ (where c and $c_s$ are the monomers and excess salt concentration, respectively) is the concentration of all free monovalent ions, and b=10.2 Å (for HA) is the distance between charges along the chain. The case of HA is interesting because $L_0$ (cf helical structure) is not too large (contrary to DNA with $L_0$~500 Å), such that $L_e$ is more than a small perturbation; also all quantities still lie within the accessible size range of usual SANS.

In this letter, we look at hyaluronan shape and interactions when embedded in a mixture with another negatively charged polyelectrolyte chain, flexible, using Small-Angle Neutron Scattering (SANS). The trick is here to combine SANS with isotopic labelling. In this purpose we use flexible d-sodium polystyrene sulfonate (NaPSS-d), which is deuterated in order to make its contribution to the scattering negligible when dissolved in $D_2O$ (more precisely at 98% $D_2O$ and 2% $H_2O$; cf scattering lengths: $\rho_{D2O}$=6.39×10$^{10}$ cm$^{-2}$, $\rho_{NaPSS-d}$=6.26×10$^{10}$ cm$^{-2}$, $\rho_{HA}$=2.335×10$^{10}$ cm$^{-2}$). The SANS signal is arising only from hydrogenated HA. We used two weight-average molecular weights (160K and 700K, determined using light scattering experiments [7]) of hydrogenated bacterial hyaluronan HA (poly[(1→3)-β-D-GlcNAc-(1→4)-β-D-GlcA]; monomer mass: 400 g/mol), which is produced and carefully purified under the Na salt form by Soliance (Pomacle, France). 80K deuterated NaPSS was purchased from Polymer Standards Service (Mainz, Germany). Dilute equimolar mixtures (1.8 g/l 160K HA + 0.96 g/l NaPSS-d; and 1 g/l 700K HA + 0.53 g/l NaPSS-d) and mixtures with similar concentrations (1 g/l 700K HA + 1 g/l NaPSS-d) were prepared in $D_2O$. External salt NaCl (0, 10$^{-3}$, 8×10$^{-3}$, and 0.2 M) was progressively added to increase the solution ionic strength. SANS measurements were carried out at T=20 °C on spectrometers D11 (ILL, Grenoble) and PACE (LLB, Saclay) with configurations allowing a large range of q. The final spectra are given in absolute units of cross section (cm$^{-1}$) following the standard procedures [7]. Also, the incoherent background of polyelectrolytes was subtracted. In dilute and high ionic strength solutions, it is classical to assume that the scattering is arising from isolated chains with $I(q)(cm^{-1}) \sim (\Delta\rho)^2 \phi V_p P(q)$, where $(\Delta\rho)^2$ is the contrast factor, $\phi$ the monomer volume fraction, and $V_p$ the chain volume.

To interpret the scattering, it is very useful to have a correct expression of the scattering of an individual chain (form factor), $P(q)$. We use an expression [11] based on the expressions derived by Burchard and Kajiwara for rodlike structures [12] in which the form factor calculated by Sharp and Bloomfield [13] for finite wormlike chains of contour length $L_c$ is used at low q.

$$P(q) = \left(\frac{2[\exp(-x)+x-1]}{x^2} + \left[\frac{4}{15} + \frac{7}{15x} - \left(\frac{11}{15} + \frac{7}{15x}\right)\exp(-x)\right]\frac{2L_T}{L_c}\right) \times \exp\left[-\left(\frac{2qL_T}{\alpha}\right)^\beta\right]$$
$$+ \left(\frac{1}{2L_c L_T q^2} + \frac{\pi}{qL_c}\right) \times \left(1 - \exp\left[-\left(\frac{2qL_T}{\alpha}\right)^\beta\right]\right) \quad (2)$$

with $x = L_c L_T q^2/3$ and valid for $L_c > 4L_T$. Values of the empirical parameters $\alpha = 5.53$ and $\beta = 5.33$, which contribute to the balance between the low-q and large-q terms, have been optimized in reference 11. To cope with the rather large index of polydispersity of HA (1.3), the contour length was taken as a fitting parameter. Eq 2 does not account for excluded volume interactions, which are negligible in the SANS q-range where $L_T$ is determined [11].

**First consider simple HA solutions** (without NaPSS-d): Figure 1 displays the scattering patterns (low-q light scattering is also presented after rescaling to SANS by the ratio between the contrasts of the two techniques [7]) for two salinities. At high I ([NaCl]=0.2 M, $\kappa^{-1}$=6.8 Å ), the scattering curve for a polymer concentration c=1.8 g/l, that is below the overlap concentration, exhibits three domains: (i) a low-q smooth variation with measurable zero-q limit I(q=0), analogous to a Guinier regime, leading respectively to a radius of gyration equal to 350 Å and to a weight-average molecular weight, $M_w$ = 160K; (ii) an intermediate polymer coil regime in which the q dependence of the scattered intensity can be described by a power law with exponent close to -2, like in Gaussian coils, and finally (iii) a $q^{-1}$ domain at higher q characteristic of a rigid rodlike behaviour for distances smaller than $L_T$. This ensemble of variations can be considered as a form factor and fitted satisfactorily by the wormlike chain model with no excluded volume interactions (Fig. 1), yielding $L_T$=43 Å (approximately equal to the intrinsic one) and $L_c$=3880 Å corresponding to the theoretical value of a single 160K HA chain. Ionic strength screens chain interactions.

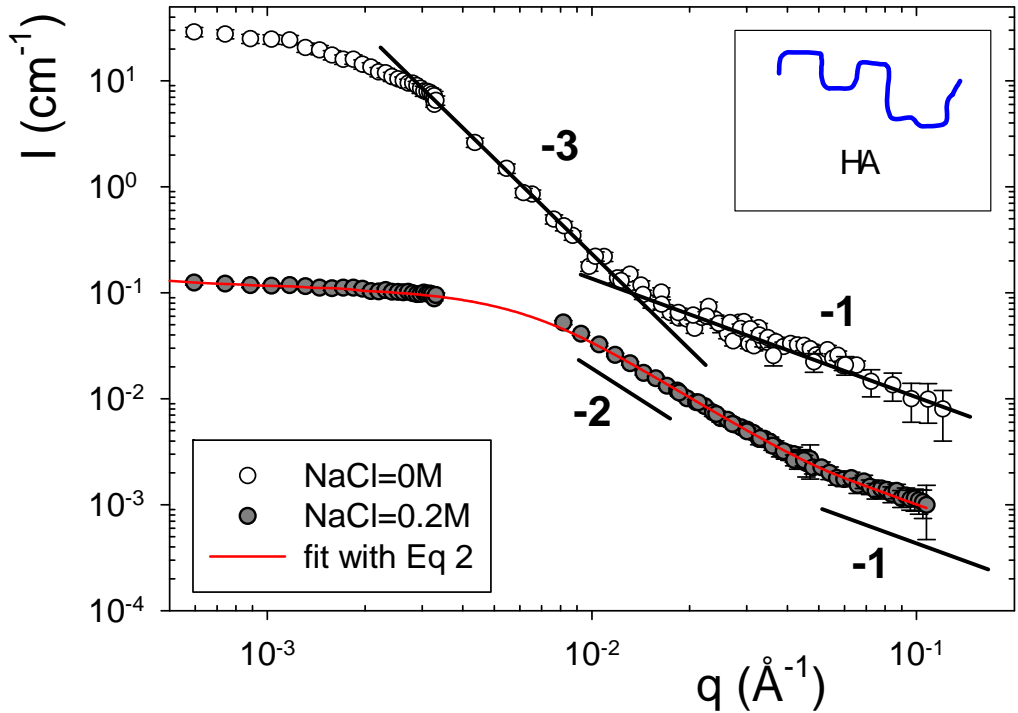

FIG. 1. Scattering for simple 160K HA solutions at c=1.8 g/l. For clarity, the curve at 0M NaCl ($\kappa^{-1}$=64 Å) was shifted by a factor 10.

At low I ([NaCl]=0 M, $\kappa^{-1}$=64 Å), screening is weak. What is expected is that the absence of screening induces active repulsions between chains, which should lead to a privileged distance, in other words to a maximum in the scattering, the so-called "polyelectrolyte peak". This is not observed: instead, we see now, for decreasing q, a transition from a $q^{-1}$ to a $q^{-3}$ dependence (instead of $q^{-1}$ to $q^{-2}$ at high I). This steep upturn probably overlaps and hides the polyelectrolyte peak (which is usually weak for some other natural polymers). The strong $q^{-3}$ scattering is usually attributed to chain aggregates. This is confirmed by the low-q light scattering, which yields, through a Guinier law, a very large radius of gyration of 1040 Å and a molecular mass of $2.4 \times 10^6$ g/mol, around 15 times the mass of a single HA chain. Finally, dynamic light scattering experiments performed on the same solutions also confirm the presence of aggregates: they display the characteristic slow polyelectrolyte mode [7] usually observed for flexible polyelectrolytes at low ionic strength [6]. Let us now look at the $q^{-1}$ regime. First, at the largest q, it overlaps perfectly with the one measured at high I (overlap is not seen on Figure 1 due to the shift used for clarity): aggregate scattering thus appears as negligible in this regime. The scattering can thus be fitted by the expression of Burchard and Kajiwara [12], using the same mass per unit length than that for single HA chain, and the same contour length $L_c$=3800 Å. The $q^{-1}$ domain is more extended towards small q, meaning that the persistence length of HA has increased. However, the $q^{-1}$-$q^{-2}$ rod-to-coil transition is masked by the aggregate signal, and it is not possible to deduce $L_T$ directly. Note that we need strong screening to prevent aggregation: $\kappa^{-1}$ must be smaller than 30 Å (for which it is still observed).

**In mixtures of HA with deuterated NaPSS** (where we "see" only HA), the effect of ionic strength is profoundly different. Both [NaCl]=0.2 M and 0 M lead now to the **same shape** of the HA scattering. No low-q light scattering data are obtained since NaPSS-d gives non-negligible signal in light scattering and the isolated HA intensity is not available. In Figure 2 we have selected the example of 160K HA (c=1.8 g/l) with deuterated NaPSS (c=0.96 g/l). A crossover between the $q^{-2}$ coil regime (instead of $q^{-3}$) and the $q^{-1}$ rod-like regime is observed even for very low ionic strength solutions (i.e., for $\kappa^{-1}$ as large as 66 Å). So, surprisingly, **even at low ionic strength the low-q aggregation behaviour is not** **observed in mixtures**. We are thus tempted to fit the scattering to the wormlike chain form factor Eq. 2: for the mixture at [NaCl] = 0 M (Fig. 2), we get a mass of 160K, and $L_c$=4000 Å, just like for single chain: the scattering is identical to the form factor. We thus access to $L_T$ at low I, with a value of 71.1 Å. For the same mixture with added salt (high I), we obtain the same $M_w$ and $L_c$, and $L_T$=43.8 Å, i.e., values for a single neutral HA chain. Fits for all polyelectrolyte and salt concentrations investigated, and two molar masses, lead to: M=160±20K and $L_c$= 4000±200 Å, for 160K HA-NaPSS-d mixtures, and M=700±70K and $L_c$=17500±2000 Å, for 700K HA-NaPSS-d mixtures. The $q^{-1}$ parts are superimposed for all conditions with only a weak dependence on excluded volume interactions and polydispersity in this q-range. This is fully similar to single HA chains behaviours, but for a much larger range of I (6.7≤$\kappa^{-1}$≤66 Å) and is the main new fact of this report.

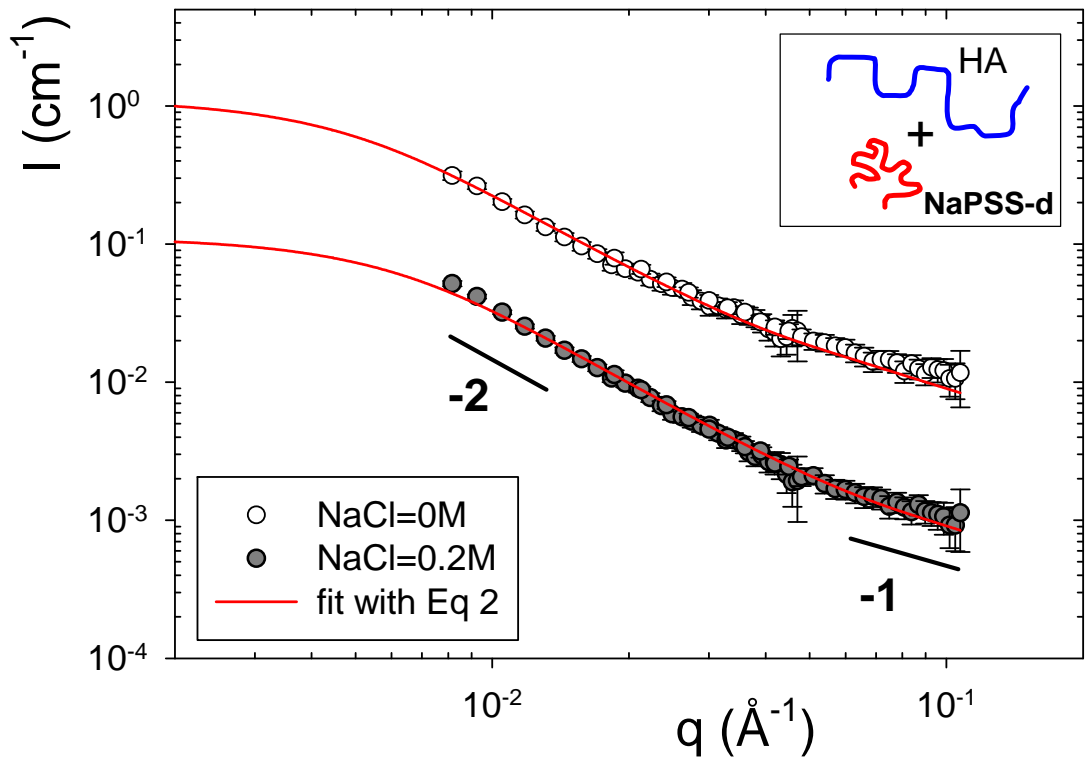

FIG. 2. Scattering for [160K HA (1.8 g/l)-80K NaPSS-d (0.96 g/l)] mixtures at low ionic strength I ($\kappa^{-1}$=55 Å, for clarity the curve has been shifted by a factor 10) and high I ($\kappa^{-1}$=6.8 Å).

This fact has an important consequence: now data sustain validity for extracting the electrostatic contribution, $L_e$, and comparing its dependence over ionic strength with theory. To get $L_e$, we need to subtract from $L_T$ the value of $L_0$: Since $L_0$ is not the focus of the paper, we get its value by subtracting from $L_T$ measured at high I ([NaCl]=0.2 M, $\kappa^{-1}$=6.8 Å), the electrostatic contribution at this I as predicted by OSF theory (Eq. 1), $L_e$=0.8 Å: the real value is bound to be as small anyway. The result, $L_e = L_T - L_{T-0.2M} + 0.8$, is shown in Figure 3 as a function of $\kappa^{-1}$. We observe that data lies upon to the line from OSF Eq. 1 **without adjusting factor**. This brings the second important result to this work, since to our knowledge; it is the first check of OSF theory with semiflexible polyelectrolyte covalent chains through direct measurement of their conformation. Former direct SANS measurements concerned flexible chains, NaPSS, and lead to a variation of $L_e$ closer to $\kappa^{-1}$ [14]. We previously attempted to follow such variation in hyaluronan only solutions, but had only indirect access to $L_e$ at low I because of aggregates scattering [7].

In summary, when HA semi-rigid chains are mixed with flexible polyelectrolytes bearing charges of the same sign at low ionic strength, an unambiguous set of experimental results shows that: (i) we neither observe an aggregation behaviour (ii) nor a polyelectrolyte peak; (iii) instead we observe a crossover between a $q^{-2}$ and a $q^{-1}$ regime, and finally (iv) a $L_e \sim \kappa^{-2}$ variation. These two findings support each other through two lines of view: (i) the well known theory of Odijk finds here a direct check. (ii) The experimental knowledge of charged chain solutions and mixtures is enriched: aggregation and interactions, generally observed for single species, vanish here when a second charged chain of same sign is added.

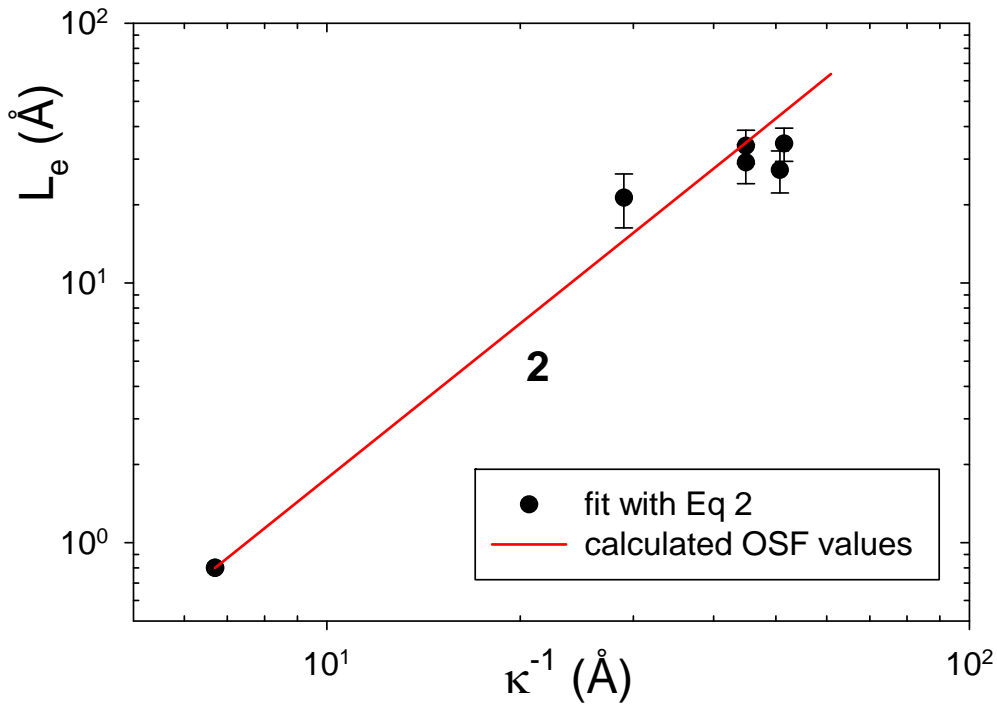

FIG. 3. Dependence on $\kappa^{-1}$ of the HA electrostatic persistence length in the different mixtures. The red line represents the actual OSF equation (Eq 1) without adjusting factor.

The origin of such aggregates in low ionic strength polyelectrolyte solutions (or, in terms of dynamic scattering, of slow modes) is not well understood. Since it occurs when long range electrostatic interactions are possible (low I), it has been proposed that polyions and counterions create a co-organized structure [8]. It occurs not only for HA only solutions as shown here, but also for flexible chains like NaPSS, which has been studied in detail [6]. But the two species have strongly different parameters: (i) NaPSS has a short intrinsic persistence length, $L_{0,PSSNa}$ =9.5 Å [14], whereas, for HA, $L_0 \geq 50$ Å. (ii) The distance between charges along the chain is b = 2.45 Å for NaPSS, which is much shorter than the Bjerrum length $l_B$ and thus induces counterion condensation as accounted by Manning theory [15]. For HA, b=10.2 Å, a much larger value, which does not induce condensation. Finally, non-electrostatic interaction between different species may play a role. Altogether, the situation is very different from a single species solution. The sharing of condensed counterions between two different species of polyelectrolytes characterized by strongly different structural parameters is difficult. It is thus possible that attractive forces between polyions mediated by this sharing become negligible, so that mixed aggregates cannot be stabilized. In other words, adding a different species creates a kind of "entropic", or "chaotropic screening". Note that mechanisms for regulating the colloidal stability have been proposed for other electrostatics systems [16]: negligibly charged colloidal microspheres, which flocculate when suspended alone in solution, undergo a stabilizing transition upon the addition of highly charged nanoparticle species that segregate to regions near microspheres because of their repulsive electrostatic interactions. In the present case the situation is different: in particular, both polyelectrolytes are well charged, and the length scales are quite shorter. Beyond absence of aggregation, we have to account for absence of polyelectrolyte peak or any interchain HA correlation (the zero q limit stays proportional to the single chain molar mass): this also could be linked with larger entropy in our mixture.

*Corresponding authors. Email address: eric.buhler@univ-paris-diderot.fr; schweins@ill.eu